\documentclass[12pt]{article}
\usepackage{amssymb,amsmath,amsfonts,amsthm,graphicx}
\def\be{\begin{equation}}
\def\ee{\end{equation}}
\def\sp{\sigma_+}
\def\Sp{\Sigma_+}
\def\R{{}^3\!R}
\def\S{{}^3\!S_+}
\def\calS{\mathcal{S}_+}

\def\E{\mathcal{E}}

\begin{document}

\begin{center}
{\Large\bf Demonstration of the spike phenomenon using the LTB models}
\vspace{.3in}
\\{\bf A A Coley},
\\Department of Mathematics \& Statistics, Dalhousie University,\\
Halifax, Nova Scotia, Canada B3H 3J5
\\Email: aac@mathstat.dal.ca
\vspace{.1in}
\\{\bf W C Lim},
\\Department of Mathematics, University of Waikato, Private Bag 3105, Hamilton 3240, New Zealand
\\Email: wclim@waikato.ac.nz
\vspace{.1in}
\vspace{0.2in}

\end{center}

[PACS: 98.80.Jk]

\begin{abstract}
  
We demonstrate the occurrence of permanent spikes using the Lema\^{\i}tre-Tolman-Bondi models, chosen because the solutions are exact
and can be analyzed by qualitative dynamical systems methods. Three examples are given and illustrated numerically. The third example
demonstrates that spikes can form directly in the matter density, as opposed to indirectly in previous studies
of spikes in the Kasner regime. Spikes provide an alternative general relativistic mechanism for generating
exceptionally large structures observed in the Universe.

\end{abstract}

\newpage
\section{Introduction}

Spikes, originally found in the context of vacuum orthogonally transitive $G_2$ models~\cite{art:BergerMoncrief1993,art:RendallWeaver2001,art:Lim2008,art:Limetal2009}, describe a dynamic and spatially inhomogeneous gravitational 
distortion. Berger and Moncrief first discovered spikes in their numerical simulations~\cite{art:BergerMoncrief1993}. Rendall and Weaver~\cite{art:RendallWeaver2001} discovered a composition of two transformations that can map 
spike-free solutions to solutions with spikes. Using the Rendall-Weaver transformation, Lim discovered an exact solution for spikes~\cite{art:Lim2008}, which suggests a numerical zooming technique that vastly improves the numerical 
accuracy of spike simulations~\cite{art:Limetal2009}.

In \cite{art:ColeyLim2012}
we explicitly showed that spikes naturally occur in a class of
non-vacuum $G_2$ models and, due to
gravitational instability, leave small residual imprints on matter in the
form of matter perturbations.
We have been particularly interested in recurring and complete
spikes formed in the oscillatory regime (or recurring spikes for short)~\cite{art:Lim2008,art:Limetal2009}, and their
imprint on matter and structure formation.
We have obtained further numerical evidence for the existence of spikes and general relativistic matter perturbations~\cite{art:LimColey2014}, which
support the results of \cite{art:ColeyLim2012}.

The residual matter overdensities from recurring spikes
form on surfaces, not at isolated points.
In $G_2$ models these spike surfaces are parallel and do not intersect.
In general spacetimes however, two spikes surfaces may intersect along a curve,
and this curve may intersect with a third spike surface at a point, leading to matter inhomogeneities forming on a web of surfaces, curves and points.
\footnote{This would be consistent with Zeldovich's idea of matter initially forming in a network of pancakes, filaments, and clumps~\cite{art:ShandarinZeldovich1989},
before finally becoming concentrated in separate clumps.}
Indeed, there are tantalising hints (from dynamical and numerical analyses) that
filamentary structures and voids would occur naturally in this scenario.
We have speculated  \cite{art:ColeyLim2012} as to whether these
recurring spikes might be an alternative to the inflationary mechanism for
generating matter perturbations and thus act as seeds for the subsequent formation of large scale structure.
Superficially, at least, there are some
similarities with perturbations and structure formation created in cosmic string models.
The inhomogeneities occur on closed circles or infinite lines \cite{art:ColeyLim2012}, 
similar to what happens in the case of
topological defects,
and it is expected that a mechanism akin to the Kibble causality mechanism will
ensure that ``defects'' form and persist to the present time. This was discussed in~\cite{art:LimColey2014}.

However, with a tilted fluid, the tilt provides another mechanism in generating matter
inhomogeneities  due to the non-negligible 
divergence term caused by the instability in the tilt.
In~\cite{art:LimColey2014} we investigated the evolution equations of 
the plane-symmetric orthogonally transitive $G_2$ model with a perfect fluid.
We numerically investigated the effect of a sign change in the tilt, which 
leads to both overdensities and underdensities through a large divergence term.
We then determined that when there is a complete spike but the tilt does not change sign,
the spikes leave a very small imprint on the matter density.
Finally, when the spike and the sign change in the tilt coincide, we found that
the spike drives the divergence term towards making overdensities as the universe expands.
We concluded that it is the tilt instability that plays the primary role in the formation of matter inhomogeneities in these models, 
but it does so by creating
local overdensities and underdensities without particular preference for either one.
On the other hand, the spike mechanism plays a secondary role in generating matter
inhomogeneities -- it drives the divergence term towards making local overdensities.
A negative divergence term creates a local overdensity.
While the divergence term itself does not prefer one sign or another, 
spikes drive the divergence term towards negative value, creating a web of local overdensities.

To date, we have
concentrated on how spikes generate matter overdensities in a radiation fluid in a
special class of inhomogeneous models. These spikes occur in the initial
oscillatory regime of general cosmological models.  
Now we want to explore spikes that occur in other regimes.
In order to find them, it is essential to understand why spikes form.

The dynamical reason for spike formation is that the initial data straddle the stable manifold of a saddle point. 
As the solution evolves, the data on either side of the stable manifold first approach the saddle point and then leave the neighbourhood of the saddle point and diverge.
The one datum that lies on the stable manifold also approaches the saddle point, but never leaves.
This heuristic argument holds as long as spatial derivative terms have negligible effect. In the case of recurring spikes, the datum on the stable manifold does leave, due to
large spatial derivative terms.

\begin{figure}[t!]
\begin{center}
    \includegraphics[width=8.5cm]{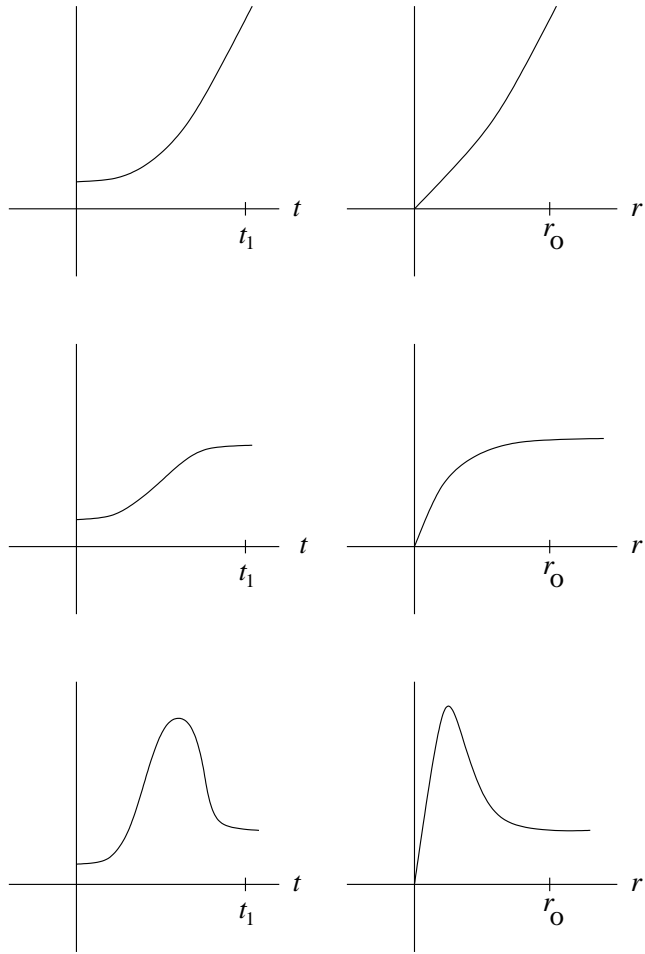}
    \caption{The essence of the straddling mechanism. The left column shows the time evolution of the unstable state space variable $M_1(t,r)$ along some spatial point $r=r_0$. The right column shows the corresponding snapshot of 
$M_1$ at a later time $t=t_1$. Three typical cases are shown. In the first case, $M_1$ blows up. In the second case, $M_1$ tends to a finite value monotonically. In the third case, $M_1$ tends to a finite value but not 
monotonically. The corresponding snapshots show a steep gradient in the first case, a step-like profile in the second case, and a spiky profile in third case.}
    \label{fig:straddle_mech}
\end{center}
\end{figure}

We now briefly illustrate the essence of the straddling mechanism with some illustrations. Suppose that the stable manifold is described by $M_1=0$, where $M_1(t,r)$ is a state space variable (to be defined below). We choose 
initial data at some $t=0$, such that $M_1=0$ at some $r=r_\text{spike}$. For example, we can choose $M_1(0,r) = r$. Then the initial data straddle the stable manifold at $r=r_\text{spike}$.
The evolution of $M_1$ along $r\neq r_\text{spike}$ can be anything. Three typical cases are illustrated in Figure~\ref{fig:straddle_mech}, which shows the time evolution of $M_1$ along some $r=r_0$, and the snapshot of $M_1$ at 
some time $t=t_1$. 
In the first case, $M_1$ blows up. The corresponding snapshot at $t=t_1$ shows a steep gradient in $M_1$. In the second case, $M_1$ grows monotonically and tends to a constant value. The corresponding snapshot at $t=t_1$ 
shows a step-like profile for $M_1$. In the third case, $M_1$ tends to a constant value but the approach is not monotone. The corresponding snapshot at $t=t_1$ shows a spiky profile for $M_1$.
The spikes in the oscillatory regime~\cite{art:Lim2008} look like the third case. The spikes that we will demonstrate in this paper look like the first and second cases. In all cases, the underlying mechanism is the same -- the 
straddling of the stable manifold. For an earlier discussion of the straddling mechanism, see~\cite[Section 6.3]{thesis:Lim2004}.
In general, the straddling of the stable manifold is a sufficient requirement for spike formation, but not necessary. The data along neighbouring worldlines need only diverge from each other
to give a spiky profile at later times. And the idea of saddle points can be generalized to include unstable subspaces. In general, such spikes may occur on sets of different dimensions (points, curves or surfaces).

The Einstein field equations of general relativity admits many self-similar solutions~\cite{art:CarrColey1999}, 
which are represented as equilibrium points in the reduced expansion-normalized state space~\cite{book:WainwrightEllis1997}. 
Many of the equilibrium points are saddle points.
It is thus natural to expect that spikes occur around these saddle points.
We shall therefore look at some of these saddle points, and construct initial data that form spikes around the saddle points.

As a first step in this direction, we shall investigate a simple model.
We shall choose the Lema\^{\i}tre-Tolman-Bondi (LTB) models (spherically symmetric dust models) to demonstrate spikes, because the LTB models are described by exact solutions. 
This makes the demonstration of spikes easier to establish.
The models studied are meant as a demonstration of the principle of spike formation in inhomogeneous spacetimes, and are not necessarily intended as physical models concordant with observational data.

\section{The LTB models}

The line element of the LTB models is
\be
        ds^2 = - dt^2 + \frac{ [R'(t,r)]^2 }{ 1+2E(r) } dr^2 + R(t,r)^2 [ d\theta^2 + \sin^2\theta d\phi^2 ],
\ee
where the dust is assumed to be co-moving. An overdot denotes a $t$-derivative, and a prime denotes an $r$-derivative.

There is only one time-dependent metric variable, $R(t,r)$, and its evolution equation is
\be
\label{master_eq}
        \dot{R}^2 = \frac{2M}{R} + 2E,
\ee
where $E(r)$ and $M(r)$ are $r$-dependent functions to be specified.

In fact, for the case $E\neq0$, $R(t,r)$ can be written down explicitly in terms of $\eta(t,r)$, and $\eta(t,r)$ implicitly in terms of $t$ and $r$.
For the case $E=0$, $R(t,r)$ can be written down explicitly in terms of $t$ and $r$.
\begin{alignat}{3}
E&>0:\ &	R(t,r) &= \frac{M}{2E}(\cosh \eta -1),	&	\sinh \eta-\eta &= \frac{(2E)^{3/2}}{M}(t-t_B(r))
\\
E&=0:\ &	R(t,r) &= \left[ \frac92 M (t-t_B(r))^2 \right]^{1/3}, &&
\\
E&<0:\ &	R(t,r) &= \frac{M}{(-2E)}(1-\cos \eta), &	\eta- \sin \eta &= \frac{(-2E)^{3/2}}{M}(t-t_B(r)).
\end{alignat}
Here $t_B(r)$ is the bang time, the time at which the solution is singular. See~\cite{book:Bolejkoetal2010} for more background on LTB models.

The LTB models are simple in the sense that all characteristic speeds are zero, so that the evolution along individual worldlines does not affect other worldlines.

\section{The dynamical systems point of view}

More insights can be gained by carrying out a dynamical systems analysis. 
For an alternative dynamical systems analysis of LTB models see, for example,~\cite{art:Sussman2008,art:SussmanIzquierdo2011}.
The set of variables we shall use are the Hubble-normalized kinematic and physical variables (see~\cite{book:WainwrightEllis1997} for a comprehensive background).
First, the unnormalized variables are 
\be
	        \{ H,\ \sp,\ \R,\ \S,\ \rho \}.
\ee
These are the Hubble expansion scalar, the shear, the spatial curvature, the anisotropic spatial curvature, and the dust density.
They are given in terms of $R(t,r)$, $E(r)$ and $M(r)$ by:
\begin{align}
        H+\sp &= \frac{\dot{R}}{R}
\\ 
        H-2\sp &= \frac{\dot{R}'}{R'}
\\
        \R &= -\frac{4(ER)'}{R^2R'}
\\      
        \S &= -\frac{1}{12} \R - \frac{E}{R^2}
\\
        \rho &= \frac{2M'}{R^2R'}.
\end{align}
The dimensionless Hubble-normalized variables are:
\be
        \Sp = \frac{\sp}{H},\quad \Omega = \frac{\rho}{3H^2},\quad \Omega_k = - \frac{\R}{6H^2},\quad \calS = \frac{\S}{3H^2}.
\ee
The Hubble-normalized Gauss constraint is
\be
	1 = \Sp^2 + \Omega_k + \Omega.
\ee
We also introduce the deceleration parameter $q$, expressed as
\be     
        q = 2 \Sp^2 + \frac12 \Omega.
\ee
For convenience, we introduce $M_1$ and $M_2$, defined as
\be
        M_1 = \frac13\Omega_k - 2 \calS,\quad M_2 = \frac13\Omega_k + \calS, 
\ee
or equivalently
\be
	M_1 = \frac{2E}{3R^2H^2},\quad M_2 = \frac{E'}{3RR'H^2}.
\ee
The reduced system of evolution equations in $\Sp$, $M_1$ and $M_2$ are then:
\begin{align}
        \partial_\tau \Sp &= (q - 2)\Sp + M_1 - M_2,
\\
        \partial_\tau M_1 &= (2q-2\Sp)M_1,
\\
	\partial_\tau M_2 &= (2q+\Sp)M_2,
\end{align}
where
\be
	q = \frac12 + \frac32 \Sp^2 - \frac12 M_1 - M_2,
\ee
and $\tau$ is related to $t$ and $r$ by
\be
        \frac{\partial t}{\partial \tau} = \frac{1}{H}.
\ee
The reduced system is three-dimensional. The auxiliary evolution equation for $\Omega$ is
\be
        \partial_\tau \Omega = (2q -1)\Omega.
\ee
The (decoupled) Hubble scalar itself has the evolution equation
\be
        \partial_\tau H = - (q+1) H.
\ee

The Weyl curvature components for spherically symmetric models are
\be
        E_{\alpha\beta} = \text{diag}(-2E_+,E_+,E_+),\quad H_{\alpha\beta}=0.
\ee
The Hubble-normalized $\E_+$ is defined as
\be
        \E_+ = \frac{E_+}{3H^2},
\ee
and is given by
\be
        \E_+ = \frac13\Sp + \frac13 \Sp^2 + \calS.  
\ee

Finding the equilibrium points is easy. There are six of them:

\begin{tabular}{lccc|cccc|c}
Name		&	$\Sp$	& $M_1$	& $M_2$	&	$\Omega_k$	& $\calS$	& $\Omega$ 	& $\E_+$	& Eigenvalues	\\	\hline
Flat FLRW	&	0	& 0	& 0	&	0		& 0		& 1 		& 0		& $1,-\tfrac32,1$\\[1mm]
Q Kasner	&	1	& 0	& 0	&	0		& 0		& 0		& $\tfrac23$	& $3,5,2$\\[1mm]
T Kasner	&	$-1$	& 0	& 0	&	0		& 0		& 0		& 0		& $3,3,6$\\[1mm]
Milne		&	0	& $\tfrac13$ & $\tfrac13$ & 1		& 0		& 0		& 0		& $-1,-1,-1$\\[1mm]
Bianchi VI$_{-5}$ &	$-\tfrac14$ & 0	& $\tfrac{15}{32}$ & $\tfrac{15}{16}$ & $\tfrac{5}{32}$ & 0	& $\tfrac{3}{32}$	& $-\tfrac34,-\tfrac{15}{8},\tfrac34$\\[1mm]
Bianchi III	&	$\tfrac12$ & $\tfrac34$ & 0 &	$\tfrac34$ 	& $-\tfrac14$	& 0		& 0 		& $0,-\tfrac32,\tfrac32$
\end{tabular}

Three forms of the Minkowski spacetime appear as equilibrium points (T Kasner, Milne and Bianchi III).
See~\cite{book:WainwrightEllis1997} for more background on such models.
The eigenvalues are listed in the table above.
We conclude from the sign of the eigenvalues that the sources are the Q Kasner and T Kasner equilibrium points, the only stable equilibrium point is the Milne equilibrium point,
and the rest are saddle points.

\begin{figure}[t!]
\begin{center}
    \includegraphics[width=0.6\textwidth]{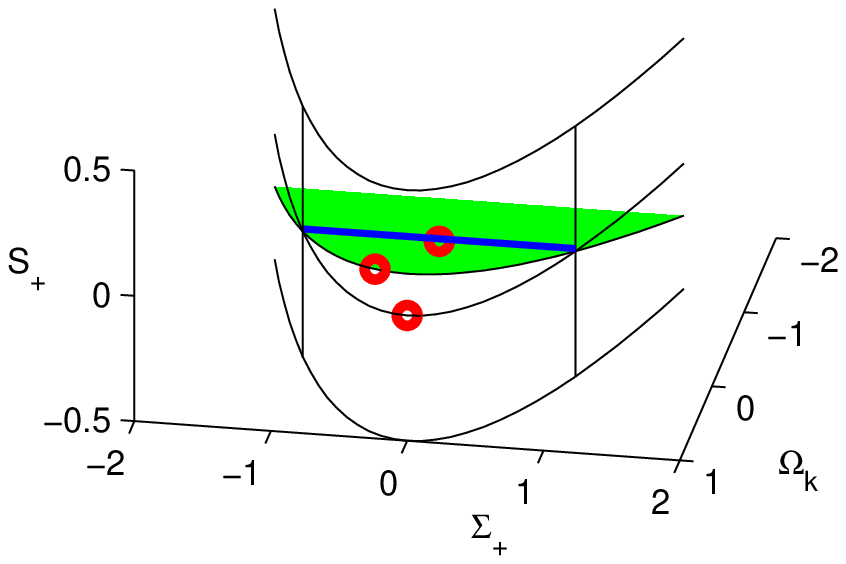}
    \caption{The Hubble-normalized state space $(\Omega_k,\Sp,\calS)$. The $E=0$ (or $M_1=0$ or $\calS=\frac16\Omega_k$) plane is shaded green. Three equilibrium points are marked with red circles:
flat FLRW at $(0,0,0)$, Milne at $(1,0,0)$, and Bianchi type VI$_{-5}$ at $(15/16,-1/4,5/32)$. The stable manifold of the flat FLRW is marked by a blue line. The stable manifold of the Bianchi type VI$_{-5}$ is the
$E=0$ plane with $\Omega_k > 0$. The parabolas mark the vacuum boundary $\Omega=0$.}
    \label{state_space}
\end{center}
\end{figure}

\begin{figure}[t!]
\begin{center}
    \includegraphics[width=0.6\textwidth]{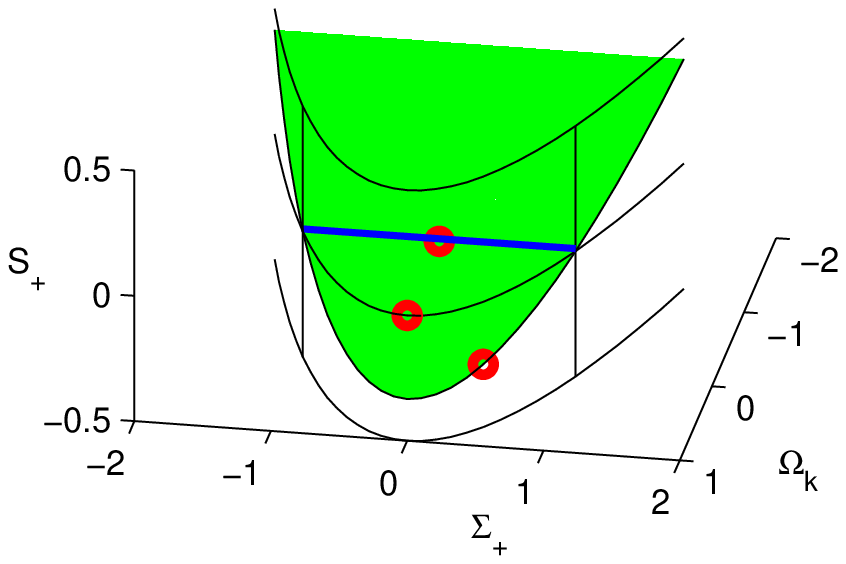}
    \caption{The Hubble-normalized state space $(\Omega_k,\Sp,\calS)$. The $E'=0$ (or $M_2=0$ or $\calS=-\frac13\Omega_k$) plane is shaded green. Three equilibrium points are marked with red circles:
flat FLRW at $(0,0,0)$, Milne at $(1,0,0)$, and Bianchi type III at $(3/4,1/2,-1/4)$. The stable manifold of the flat FLRW is marked by a blue line. The stable manifold of the Bianchi type III is the
$E'=0$ plane with $\Omega_k > 0$. The parabolas mark the vacuum boundary $\Omega=0$.}
    \label{state_spaceM2}
\end{center}
\end{figure}

Figures~\ref{state_space} and~\ref{state_spaceM2} show the Hubble-normalized state space $(\Omega_k,\Sp,\calS)$.
The existence of the two invariant sets $M_1=0$ (or $E=0$) and $M_2=0$ (or $E'=0$) divide the state space into four general subsets and five special subsets.
Both invariant sets are unstable near the flat Friedmann-Lema\^{\i}tre-Robertson-Walker (FLRW) saddle point. The future asymptotic states are listed below for initial data in each subset.

\begin{tabular}{c|ccc}
		&	$E < 0$		&	$E = 0$			&	$E >0$	\\	\hline
$E'>0$		&	recollapses	&	Bianchi VI$_{-5}$	&	Milne	\\
$E'=0$		&	recollapses	&	Flat FLRW		&	Bianchi III	\\
$E'<0$		&	recollapses	&	recollapses		&	recollapses
\end{tabular}

The four future asymptotic states for ever-expanding cases are the Milne state, the flat FLRW state, the Bianchi III state, and an 
anisotropic state best described as a Bianchi type VI$_{-5}$ plane wave state. The other cases recollapse to a singularity.

Typically, the function $E(r)$ can be chosen such that $E(r)$ has both signs for different values of $r$. Where $E(r_0)=0$, $E'(r_0)$ typically can have either sign. The only special case is $E(r_0)=E'(r_0)=0$.
Thus regions with $E(r)>0$ and $E'(r)>0$ evolve into isotropic voids; regions with $E(r)<0$ or $E'(r)<0$ recollapse to form stars (the dust equation of state is only valid up to a certain maximum density);
ever-expanding worldlines with $E(r_0)=0$ either evolve to the anisotropic Bianchi type VI$_{-5}$ state (if $E'(r_0)>0$), or to the flat FLRW state (if $E'(r_0)=0$);
and worldlines with $E(r_0)>0$ and $E'(r_0)=0$ expand to the Bianchi type III state.

Where do spikes come in? The points where $E(r)=0$ serve as a surface around which spikes can occur.\footnote{Surfaces where $E'(r)=0$ can also serve as spike surfaces. But we shall focus on the $E(r)=0$ surfaces for illustration, 
as they clearly divide the spacetime into ever-expanding and recollapsing regions.}
Because all characteristic speeds are zero, there are no spatial derivative terms to alter the dynamics of spikes. 
These spikes are permanent (as opposed to transient and recurring spikes as in~\cite{art:ColeyLim2012,art:LimColey2014}). 
The spatial profile of a spike will simply grow steeper and steeper.

\section{Qualitative dynamics} 

We will present a few numerical examples with different qualitative behaviour for illustration.
\begin{itemize}
\item	The first example assumes $E(r)$ has both signs, and demonstrates the straddling of the $E=0$ saddle surface.
\item	The second example assumes $E=0$, with $\sp=0$ at some $r=r_0$, and demonstrates the straddling of the flat FLRW saddle point itself. 
	But in this case spikes do not form. $E$ being identically zero restricts the state space to the stable manifold of the flat FLRW saddle point, so the flat FLRW point becomes the sink of this $E=0$ subspace.
\item	The third example assumes $E(r)\geq 0$, with $E=0$ as a local minimum, and demonstrates the straddling of the stable manifold of the flat FLRW saddle point.
\end{itemize}
In these examples, we shall focus on the worldlines within the specified range of $r$, and not worry about the analytical extension of the data to all $r$.

\begin{figure}[t!]
\begin{center}
    \includegraphics[width=\textwidth]{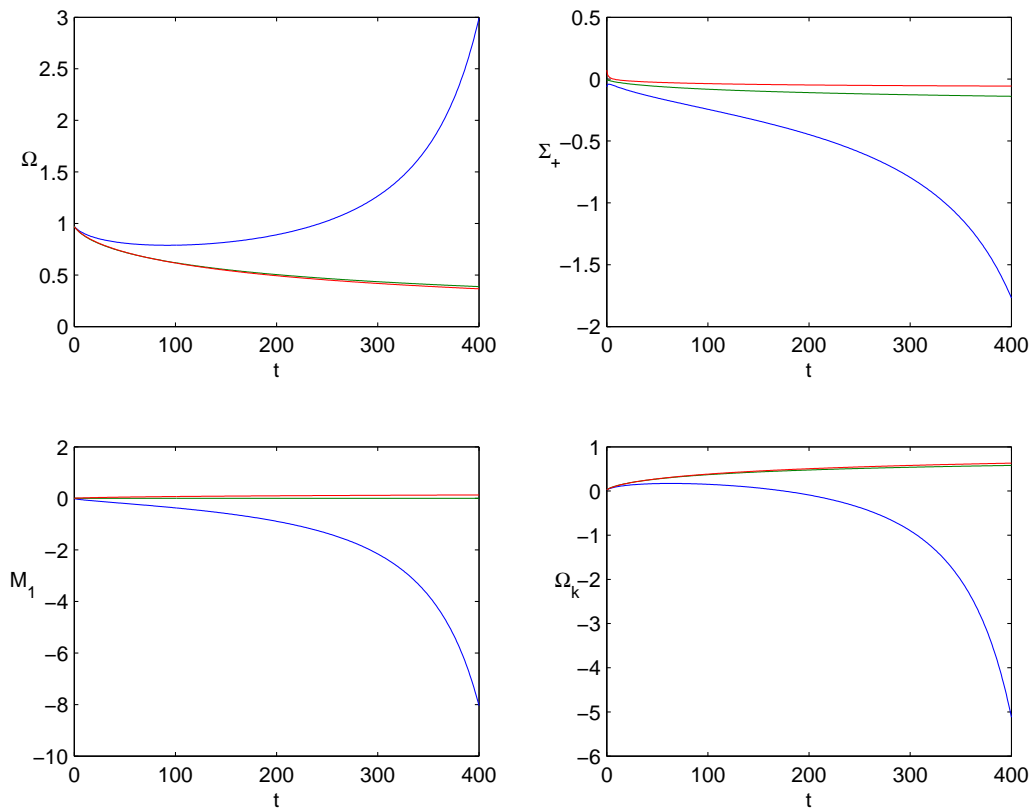}
    \caption{First example. $r=1.5$ (blue), $r=2$ (green), and $r=2.5$ (red). See text for explanation.}
    \label{first}
\end{center}
\end{figure}
\begin{figure}[t!]
\begin{center}
    \includegraphics[width=\textwidth]{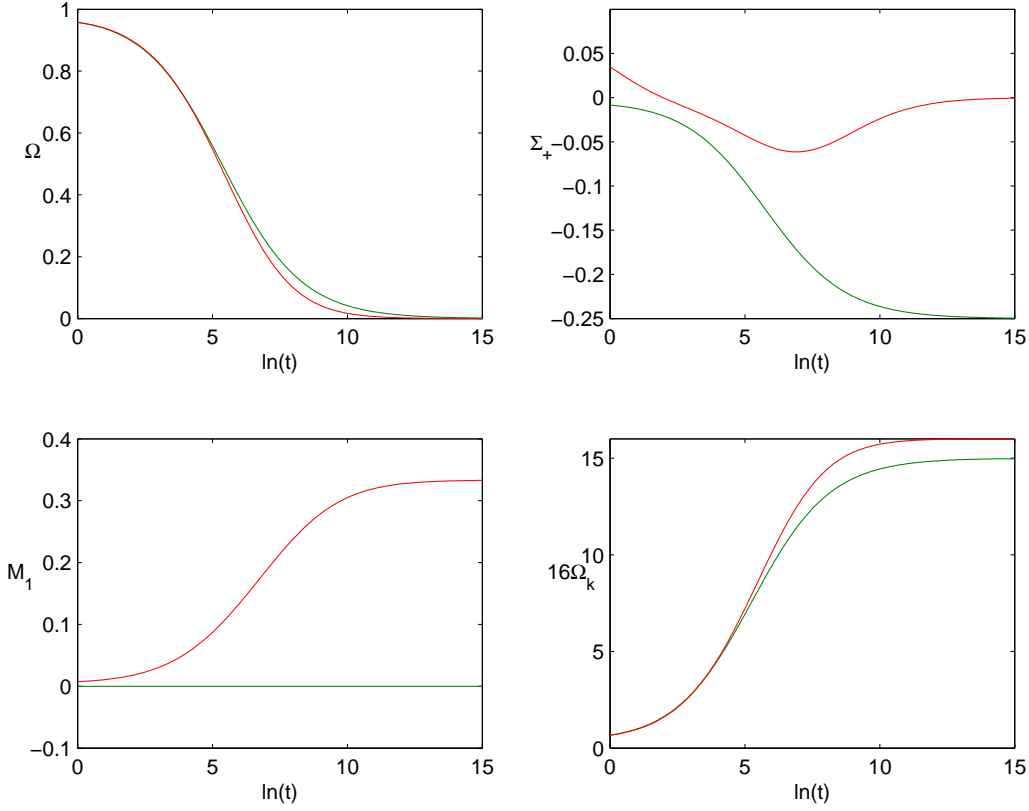}
    \caption{First example. $r=2$ (green), and $r=2.5$ (red).}
    \label{first_long}
\end{center}
\end{figure}

\begin{figure}[t!]
\begin{center}
    \includegraphics[width=\textwidth]{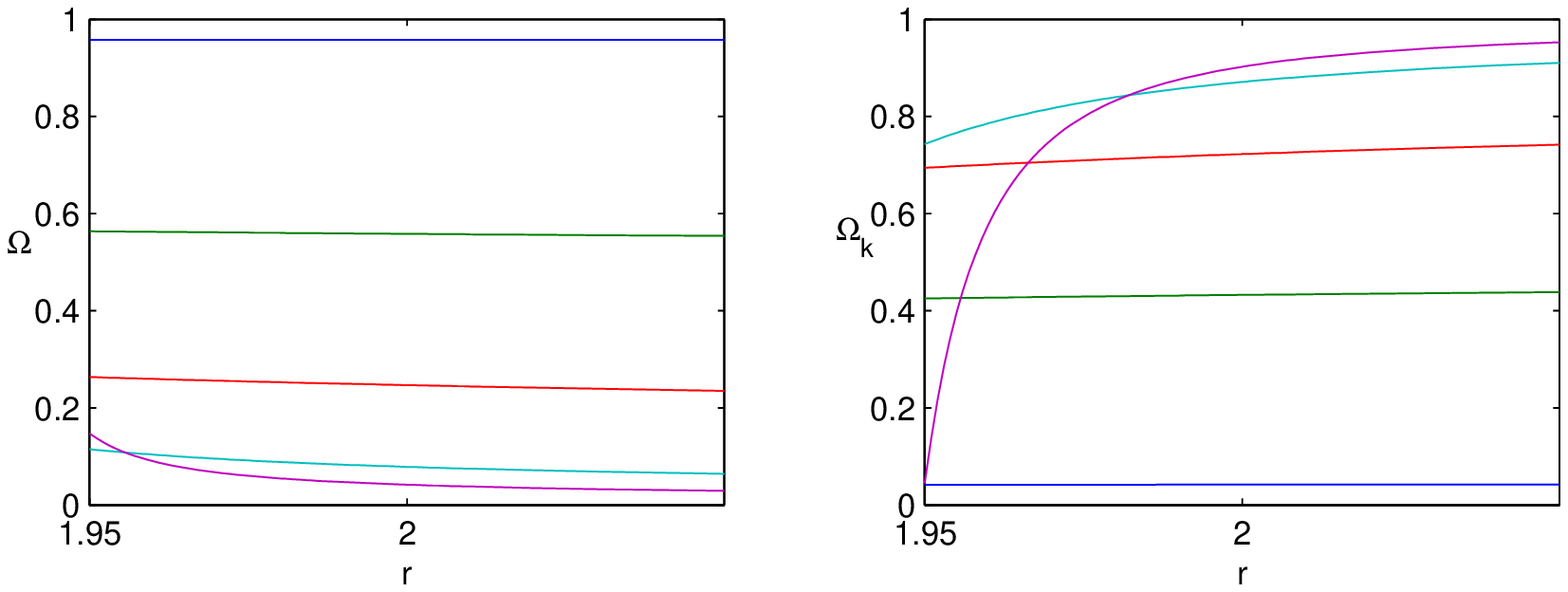}
    \caption{First example. Snapshots of $\Omega$ and $\Omega_k$ at five different times ($\ln t = 0,5,7,9,10$), showing the formation of a spike at $r=2$.}
    \label{first_detail_Alan}
\end{center}
\begin{center}
    \includegraphics[width=\textwidth]{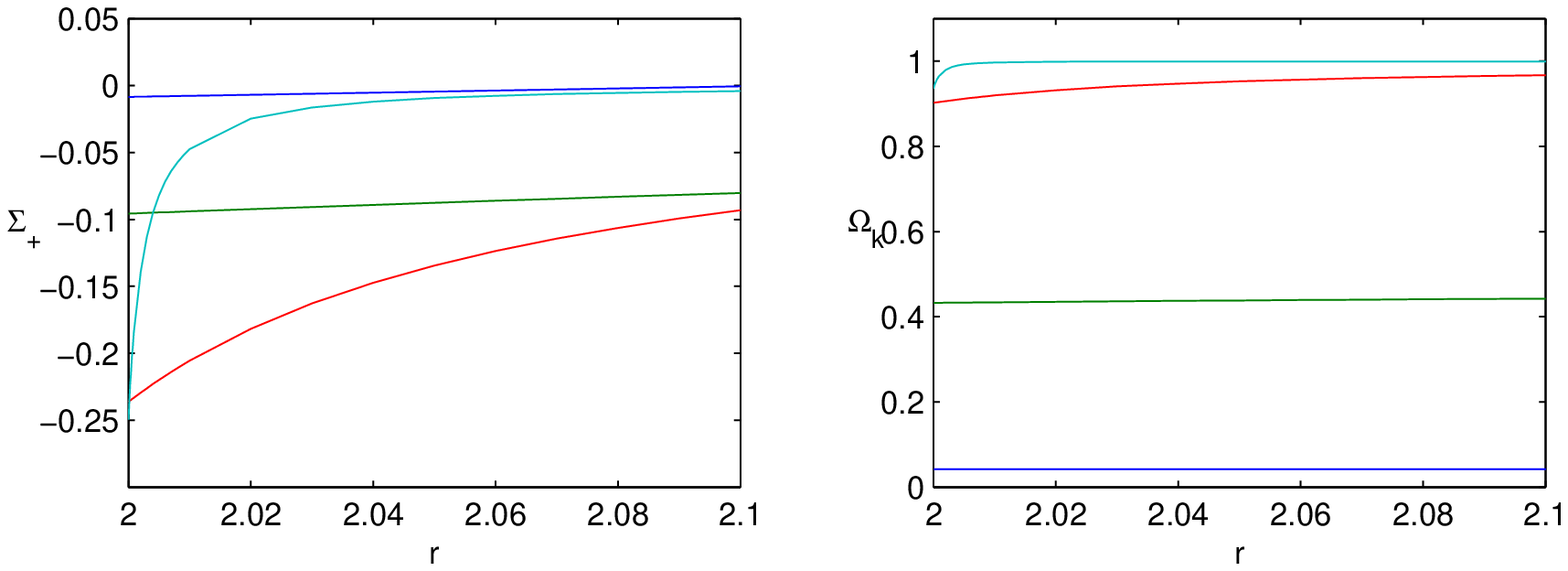}
    \caption{First example. Snapshots of $\Sp$ and $\Omega_k$ at four different times ($\ln t = 0,5,10,15$), showing the formation of a spike at $r=2$.}
    \label{first_detail}
\end{center}
\end{figure}

The first example is specified by
\be
        E(r) = 0.1(r-2),\quad  
        M(r) = r^3,\quad
        t_B(r) = -1 - 0.1 (r-2)^2.
\ee
It creates a solution straddling the $E=0$ saddle surface along the $r=2$ worldline.
Figure~\ref{first} plots $\Omega_k$, $\Sp$, $M_1$ and $\Omega_k$ along three worldlines $r=1.5$ (blue), $r=2$ (green), and $r=2.5$ (red) for $t \in [0,400]$,
showing that along $r=1.5$, the variables are blowing up due to $H$ tending to zero at the moment of maximal expansion. 
Figure~\ref{first_long} plots $\Omega$, $\Sp$, $M_1$ and $16\Omega_k$ along $r=2$ (green), and $r=2.5$ (red) for $\ln t \in [0,15]$,
showing that the $r=2$ worldline tends to the Bianchi type VI$_{-5}$ point, and the
$r=2.5$ worldline tends to the Milne point.
Figure~\ref{first_detail_Alan} plots snapshots of $\Omega$ (left panel) and $\Omega_k$ (right panel) at five different times ($\ln t = 0,5,7,9,10$), on the interval $1.95 \leq r \leq 2.05$.
Shortly after $\ln t =10$, $H$ starts to become negative in the recollapsing region $r < 2$.
Figure~\ref{first_detail} plots snapshots of $\Sp$ (left panel) and $\Omega_k$ (right panel) at four different times ($\ln t = 0,5,10,15$), showing the formation of spike at $r=2$.
The snapshots are deliberately limited to $r \geq 2$ to avoid the recollapsing region $r < 2$, where Hubble-normalized variables blow up as $H$ changes sign at the start of recollapse,
and where the physical variables blow up when $R$ tends to zero at the big crunch.

We note that while $\Omega$ tends to zero both along $r=2$ and along $r>2$, it does so at different rates. Along $r=2$, $\Omega$ tends to zero at the rate of $e^{-3/4\tau}$ or $t^{-2/3}$ in the Bianchi VI$_{-5}$ regime, while
along $r>2$, $\Omega$ tends to zero at the rate of $e^{-\tau}$ or $t^{-1}$ in the Milne regime. Along $r<2$, however, the spacetime recollapses, ending in a big crunch.
We do not mean for the dust model to be used all the way through recollapse. Realistically speaking, the dust model is valid up to a certain temperature, when stars are formed.
So the $r<2$ spacetime should be valid up to the formation of stars, while $r \geq 2$ should be valid for all times.
The $r=2$ worldline is meant to model the immediate neighbourhood of the star formation region, while $r>2$ worldlines are meant to model the voids away from this region.
The first example shows that typically, at late times, this immediate neighbourhood of the star is anisotropic (in both senses of spatial curvature and rate of expansion), and has a slightly higher matter density than the void.
The void is isotropic at late times.

\begin{figure}[t!]
\begin{center}
    \includegraphics[width=\textwidth]{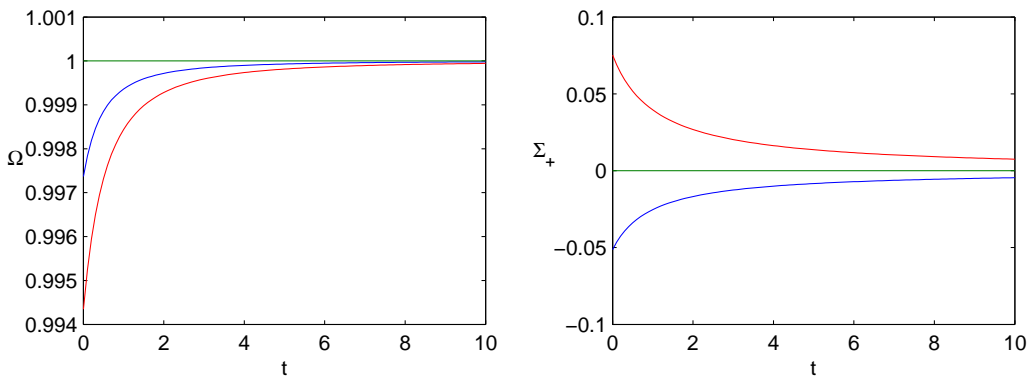}
    \caption{Second example. $r=1.5$ (blue), $r=2$ (green), and $r=2.5$ (red).}
    \label{second}
\end{center}
\end{figure}

The second example is specified by
\be
        E(r) = 0,\quad
        M(r) = r^3,\quad
        t_B(r) = -1 - 0.1 (r-2)^2.
\ee
It creates a solution straddling the flat FLRW saddle point itself along the $r=2$ worldline.
Figure~\ref{second} plots $\Omega$ and $\Sp$ along $r=1.5$ (blue), $r=2$ (green), and $r=2.5$ (red) for $t \in [0,10]$,
showing that along all orbits, $\Omega$ tends to 1 and $\Sp$ tends to 0. $\Omega_k$ is identically zero.

\begin{figure}[t!]
\begin{center}
    \includegraphics[width=\textwidth]{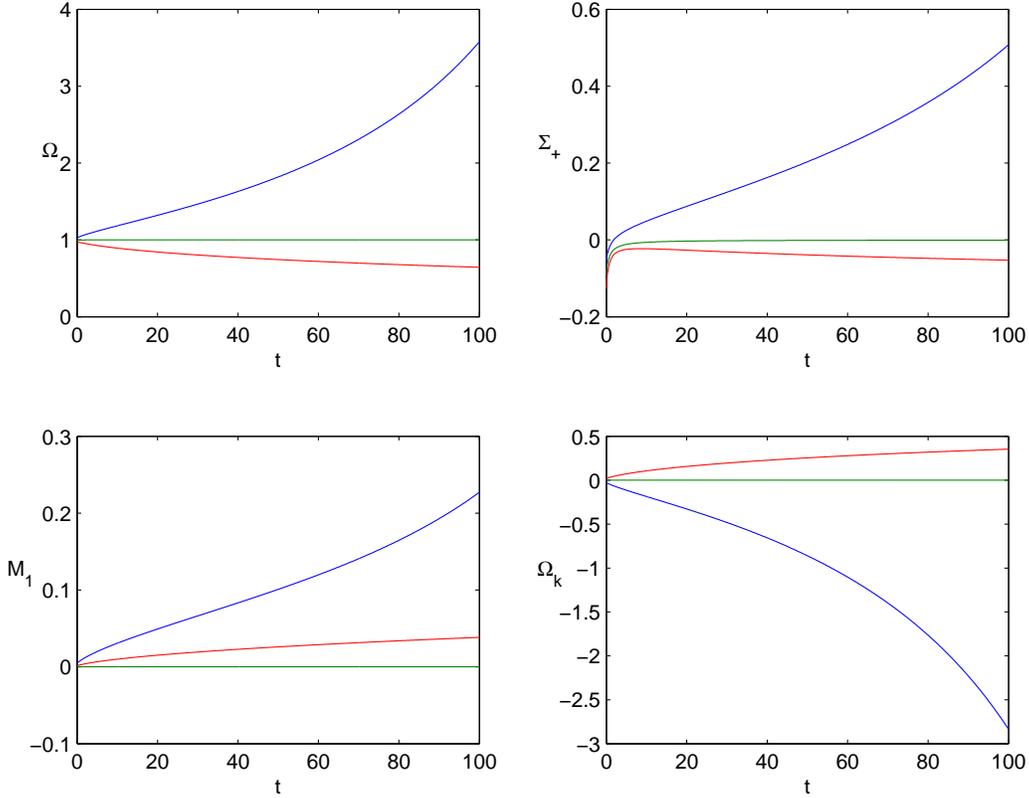}
    \caption{Third example. $r=1.5$ (blue), $r=2$ (green), and $r=2.5$ (red).}
    \label{third}
\end{center}
\end{figure}
\begin{figure}[t!]
\begin{center}
    \includegraphics[width=\textwidth]{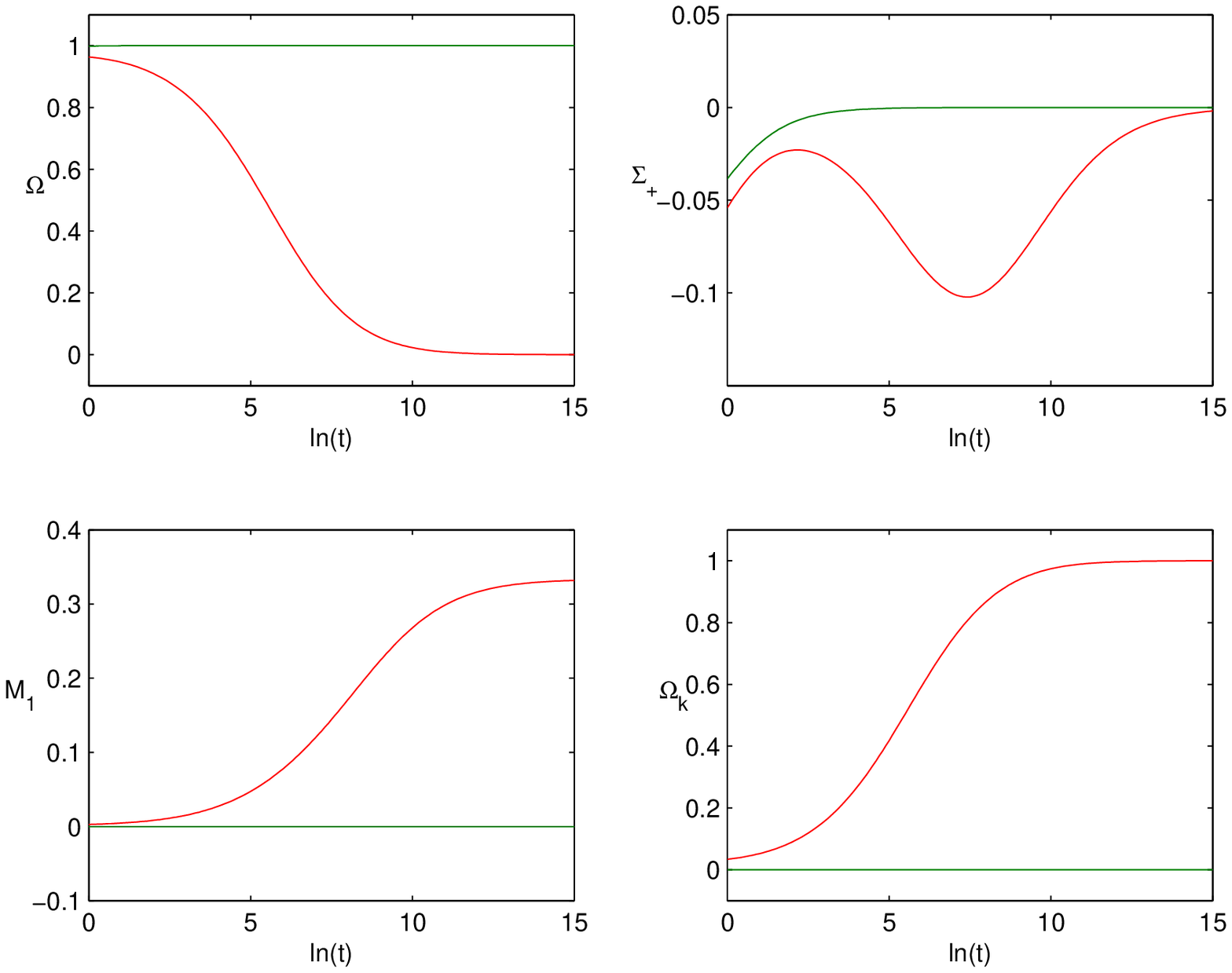}
    \caption{Third example. $r=2$ (green), and $r=2.5$ (red).}
    \label{third_long}
\end{center}
\end{figure}
\begin{figure}[t!]
\begin{center}
    \includegraphics[width=\textwidth]{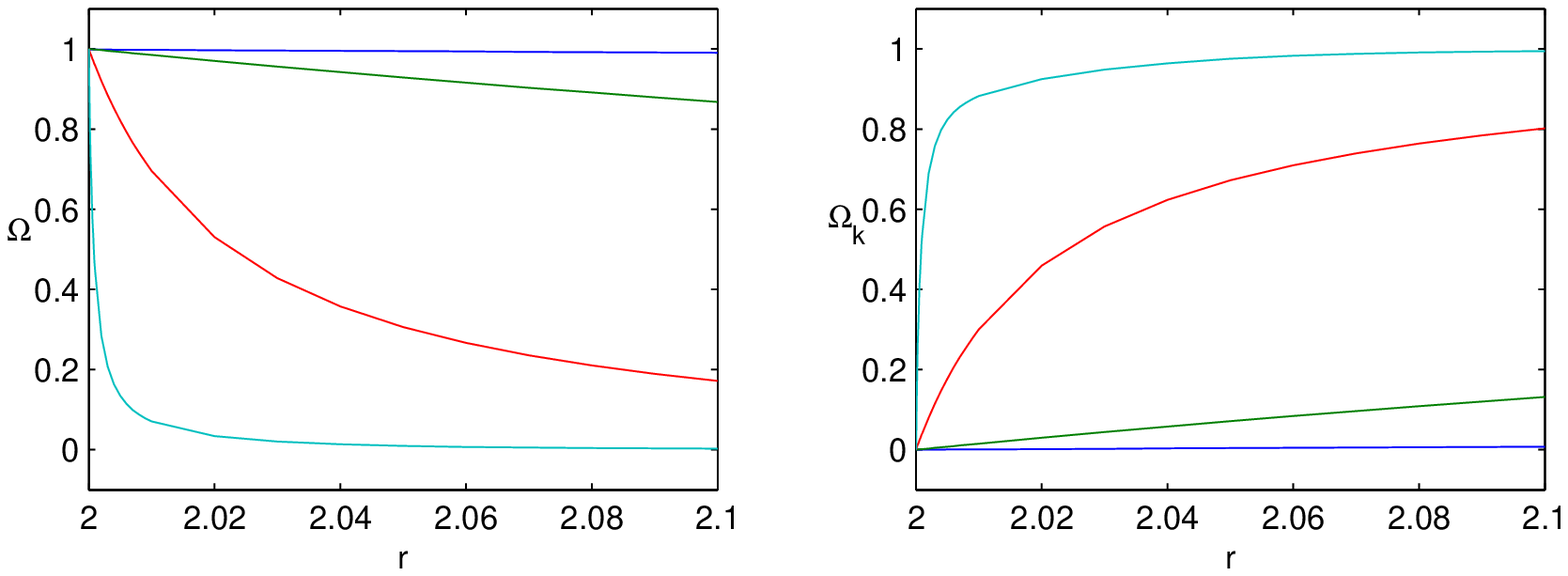}
    \caption{Third example. Snapshots of $\Omega$ (left panel) and $\Omega_k$ (right panel) at four different times ($\ln t = 0,5,10,15$), showing the formation of a spike at $r=2$.}
    \label{third_detail}
\end{center}
\end{figure}

The third example is specified by
\be
	E(r) = 0.1(r-2)^2,\quad
	M(r) = r^3,\quad
	t_B(r) = -1 + 0.1 r.
\ee
It creates a solution straddling the stable manifold of the flat FLRW saddle point, along the $r=2$ worldline.
Figure~\ref{third} plots $\Omega$, $\Sp$, $M_1$ and $\Omega_k$ along $r=1.5$ (blue), $r=2$ (green), and $r=2.5$ (red) for $t \in [0,100]$.
As in the first example, along $r=1.5$ the variables are blowing up due to $H$ tending to zero at the moment of maximal expansion.
Figure~\ref{third_long} plots $\Omega$, $\Sp$, $M_1$ and $\Omega_k$ along $r=2$ (green), and $r=2.5$ (red) for $\ln t \in [0,15]$,
showing that the $r=2$ worldline tends to the flat FLRW point, and the $r=2.5$ worldline tends to the Milne point.
Figure~\ref{third_detail} plots snapshots of $\Omega$ (left panel) and $\Omega_k$ (right panel) at four different times ($\ln t = 0,5,10,15$), showing the formation of a spike at $r=2$.

In summary, the first and the third examples exhibit qualitatively different evolution along different worldlines. A qualitative difference is generally accompanied by very large quantitative differences,
which means that a spike can generate an exceptionally large density inhomogeneity in a short time. 

Furthermore, in the third example the spike occurs in the flat FLRW regime, where $\Omega$ is non-zero.
This is qualitatively different from cases where the spike occurs in the vacuum regime, as in the first example (Bianchi VI$_{-5}$ regime) or the Kasner regime in~\cite{art:ColeyLim2012}.
The third example demonstrates that spikes can form directly in the matter density, as opposed to matter inhomogeneity forming indirectly as an imprint of spikes that form in the gravitational field only, as 
found previously in~\cite{art:ColeyLim2012}. This direct formation of spikes in the matter density creates much larger density inhomogeneities than the ``indirect mechanism''.

It is hard to make quantitative statements about the rate of expansion or recollapse.
Such statements can be made, but are restricted to the regime near equilibrium points. In particular, we are interested in the flat FLRW equilibrium point.
If one starts with an initial condition in this regime, 
then
the eigenvalues at the flat FLRW gives the following growth rates if $E(r) \neq 0$:
\be
        \Omega-1 \sim e^\tau \sim t^{2/3},\quad
        \Sp \sim e^{-\frac32 \tau} \sim t^{-1},\quad
        M_1       \sim e^\tau \sim t^{2/3}.
\ee
If $E(r)=0$, then $M_1=0$. Furthermore, if $E(r)=E'(r)=0$, then $M_1=0$ and $\Omega-1 \sim e^{-3\tau}$ or $t^{-2}$.
In summary, in the linear regime near the flat FLRW equilibrium point, $\Omega$ diverges from unity at the rate of $e^\tau$ or $t^{2/3}$.

For the nonlinear regime, one has to rely on exact solutions.
The dynamics of a solution during the time interval of interest can change considerably, evolving from one regime to another,
and as a result it does not give rise to a characteristic rate of expansion.

\section{Discussion}

The LTB models exhibit permanent spikes around the $E(r)=0$ worldlines.
Beyond LTB models, we expect similar permanent spikes to occur at the boundary between ever-expanding and recollapsing regions, and at other unstable boundaries.
A natural continuation of this investigation is to study spikes in the context of the Szekeres solutions~\cite{art:Szekeres1975} and silent universes~\cite{art:Matarreseetal1993}
(which are silent dust models like the LTB models, but admit no symmetries). 
A dynamical systems analysis of these silent models can be found in~\cite[Chapter 13]{book:WainwrightEllis1997}. The dynamical systems for LTB and Szekeres models are actually identical.
The Szekeres solutions were used as an example in~\cite{art:Limetal2006} for a different purpose, and spike formation (towards the initial singularity) was briefly mentioned at the end of Section 3.2.2 therein.

Both the incomplete spikes~\cite{art:ColeyLim2012} and the late time non-generic inhomogeneous spikes (in the third example)
 might lead to the existence of
exceptional structures on large scales. Let us discuss this in more detail.

In the standard cosmological model spatial homogeneity is only valid
on scales larger than 100-115 Mpc
\cite{art:Scrimgeouretal2012,art:Yadav2010}.
Indeed, it  only appears to be
homogeneous in some statistical sense (when averaged
on scales of
100-115 Mpc), {\footnote{Our understanding of cosmological observations is
greatly complicated by the fact that the Universe is not
completely homogeneous;
this has
led to considerable interest in the so-called averaging problem in inhomogeneous cosmology \cite{art:Wiltshire2011,art:Coley2010,art:Coleyetal2005,art:Coley2010b}.}}
 and there can be spatial inhomogeneities up to scales of at least 100 Mpc.
However, the actual Universe has a spongelike structure, dominated
by huge voids surrounded by bubble walls, and threaded by filaments, within which clusters of
galaxies are located. At the present epoch the distribution of matter is far from homogeneous
on scales less than 150-300 Mpc.
Indeed, the largest structures so far detected are limited only by the size of the surveys
that found them \cite{art:Labini2009,art:Labinietal2009}.

The Shapley supercluster has a core diameter of 40 Mpc \cite{art:Tullyetal1992}.
The Sloan Great Wall (SGW) is a large filament about
400 Mpc long \cite{art:Gottetal2005} found in the 
Sloan Digital Sky Survey (SDSS) region,
and is larger than the CfA (Great) Wall (and is
comparable with the survey size \cite{art:ShethDiaferio2011}).
Large Quasar Groups (LQGs) are the largest structures
seen in the early Universe, of characteristic size 70-350 Mpc.
A ``Huge'' Large Quasar Group (LQG) of particularly large size
(it has characteristic volume size $\sim$ 500 Mpc and longest dimension $\sim$ 1240 Mpc)
and high membership (73 quasars)
has been identified  in the DR7QSO catalogue of the SDSS
\cite{art:Clowes2012}.
The characteristic size, and especially the long dimension,
is well in excess of  the scale of homogeneity 100-115 Mpc
\cite{art:Scrimgeouretal2012,art:Yadav2010}. It appears to be
the largest feature so far seen in the early Universe.
It is also adjacent to the CCLQG, which is itself very large \cite{art:Clowes2012}.

Recent surveys suggest that a large percentage of the present volume of the
Universe is in voids of a characteristic scale 30 Mpc.
If smaller minivoids and larger supervoids
are included, then our observed Universe is presently void-dominated by volume and thus within
regions as small as 100 Mpc density contrasts of order unity are observed. 
Locally there are two enormous voids, both 35 to 70 Mpc across, associated
with the so-called velocity anomaly \cite{art:Rizzi2007}.
Also, there is evidence for a local void
on scales of 430 Mpc \cite{art:Frith2003}
and for void complexes on scales up
to 450 Mpc \cite{art:Park2012}.
{\footnote{In addition,
there  appear to be large scale
features in the CMB  \cite{art:Dunkleyetal2009,art:Ade2013,art:Urrestilla2011} and
there has been detection of cosmic flows on approximately
Gpc-scales \cite{art:Watkins2009,art:Feldman2008,art:Kashlinsky2011,art:Kashlinsky2012,art:Wiltshireetal2013}.}}

Such large inhomogeneities in the distribution
of superclusters and voids on scales of 200-300 Mpc and above, and
especially the Huge-LQG
and its proximity to the CCLQG at the
same redshift  ($\sim 1.27$) \cite{art:Clowes2012},
implies that the Universe is
perhaps not homogeneous
on these scales, and
are potentially in conflict
with  the cosmological principle and the standard concordance ($\Lambda$CDM) cosmological model
\cite{art:ShethDiaferio2011,art:Park2012}.

An alternative general relativistic spike mechanism for naturally generating (a small number of)
exceptional structures at late times (in additional to the usual
distribution of structures produced in the standard model) may resolve some tension with
cosmological observations.

\section*{Acknowledgement}

AAC is supported by NSERC.
WCL thanks Mathematical Sciences Research Institute (MSRI) Berkeley for support, where part of this research was carried out.

\bibliography{cites}

\begin{thebibliography}{10}

\bibitem{art:BergerMoncrief1993}
B.~K. Berger and V.~Moncrief,
\newblock Phys. Rev. D {\bf 48}, 4676 (1993).

\bibitem{art:RendallWeaver2001}
A.~D. Rendall and M.~Weaver,
\newblock Class. Quant. Grav. {\bf 18}, 2959 (2001), arXiv:gr-qc/0103102.

\bibitem{art:Lim2008}
W.~C. Lim,
\newblock Class. Quant. Grav. {\bf 25}, 045014 (2008), arXiv:0710.0628.

\bibitem{art:Limetal2009}
W.~C. Lim, L.~Andersson, D.~Garfinkle, and F.~Pretorius,
\newblock Phys. Rev. D {\bf 79}, 103526 (2009), arXiv:0904.1546.

\bibitem{art:ColeyLim2012}
A.~A. Coley and W.~C. Lim,
\newblock Phys. Rev. Lett. {\bf 108}, 191101 (2012), arXiv:1205.2142.

\bibitem{art:LimColey2014}
W.~C. Lim and A.~A. Coley,
\newblock Class. Quant. Grav. {\bf 31}, 015020 (2014), arXiv:1311.1857.

\bibitem{art:ShandarinZeldovich1989}
S.~F. Shandarin and Y.~B. Zeldovich,
\newblock Rev. Mod. Phys. {\bf 61}, 185 (1989).

\bibitem{thesis:Lim2004}
W.~C. Lim,
\newblock {\em The {D}ynamics of {I}nhomogeneous {C}osmologies},
\newblock PhD thesis, University of Waterloo, Canada, 2004,
  arXiv:gr-qc/0410126.

\bibitem{art:CarrColey1999}
B.~J. Carr and A.~A. Coley,
\newblock Class. Quant. Grav. {\bf 16}, R31 (1999), arXiv:gr-qc/9806048.

\bibitem{book:WainwrightEllis1997}
J.~Wainwright and G.~F.~R. Ellis,
\newblock {\em Dynamical systems in cosmology} ({C}ambridge {U}niversity
  {P}ress, Cambridge, 1997).

\bibitem{book:Bolejkoetal2010}
K.~Bolejko, A.~Krasi\'nski, C.~Hellaby, and M.-N. C\'el\'erier,
\newblock {\em Structures in the Universe by Exact Methods} (Cambridge
  University Press, Cambridge, 2010).

\bibitem{art:Sussman2008}
R.~A. Sussman,
\newblock Class. Quant. Grav. {\bf 25}, 015012 (2008), arXiv:0709.1005.

\bibitem{art:SussmanIzquierdo2011}
R.~A. Sussman and G.~Izquierdo,
\newblock Class. Quant. Grav. {\bf 28}, 045006 (2011), arXiv:1004.0773.

\bibitem{art:Szekeres1975}
P.~Szekeres,
\newblock Commun. Math. Phys. {\bf 41}, 55 (1975).

\bibitem{art:Matarreseetal1993}
S.~Matarrese, O.~Pantano, and D.~Saez,
\newblock Phys. Rev. D {\bf 47}, 1311 (1993).

\bibitem{art:Limetal2006}
W.~C. Lim, C.~Uggla, and J.~Wainwright,
\newblock Class. Quant. Grav. {\bf 23}, 2607 (2006), arXiv:gr-qc/0511139.

\bibitem{art:Scrimgeouretal2012}
M.~Scrimgeour {\em et~al.},
\newblock Mon. Not. Roy. Astr. Soc. {\bf 425}, 116 (2012), arXiv:1205.6812.

\bibitem{art:Yadav2010}
J.~K. Yadav, J.~S. Bagla, and N.~Khandai,
\newblock Mon. Not. Roy. Astr. Soc. {\bf 405}, 2009 (2010), arXiv:1001.0617.

\bibitem{art:Wiltshire2011}
D.~L. Wiltshire,
\newblock Class. Quant. Grav. {\bf 28}, 164006 (2011), arXiv:1106.1693.

\bibitem{art:Coley2010}
A.~A. Coley,
\newblock Class. Quant. Grav. {\bf 27}, 245017 (2010), arXiv:0908.4281.

\bibitem{art:Coleyetal2005}
A.~A. Coley, N.~Pelavas, and R.~M. Zalaletdinov,
\newblock Phys. Rev. Lett. {\bf 95}, 151102 (2005), arXiv:gr-qc/0504115.

\bibitem{art:Coley2010b}
A.~A. Coley,
\newblock Averaging in cosmological models,
\newblock arXiv:1001.0791, 2010.

\bibitem{art:Labini2009}
F.~S. Labini,
\newblock Super-homogeneity and inhomogeneities in the large scale matter
  distribution,
\newblock arXiv:0912.1191, 2009.

\bibitem{art:Labinietal2009}
F.~S. Labini {\em et~al.},
\newblock Astr. Astrophys. {\bf 505}, 981 (2009), arXiv:0903.0950.

\bibitem{art:Tullyetal1992}
R.~B. Tully, R.~Scaramella, G.~Vettolani, and G.~Zamorani,
\newblock Astrophys. J. {\bf 388}, 9 (1992).

\bibitem{art:Gottetal2005}
J.~R. {Gott III} {\em et~al.},
\newblock Astrophys. J. {\bf 624}, 463 (2005), arXiv:astro-ph/0310571.

\bibitem{art:ShethDiaferio2011}
R.~K. Sheth and A.~Diaferio,
\newblock Mon. Not. Roy. Astr. Soc. {\bf 417}, 2938 (2011), arXiv:2011.3378.

\bibitem{art:Clowes2012}
R.~G. Clowes {\em et~al.},
\newblock Mon. Not. Roy. Astr. Soc. {\bf 429}, 2910 (2012), arXiv:1211.6256.

\bibitem{art:Rizzi2007}
L.~Rizzi {\em et~al.},
\newblock Mon. Not. Roy. Astr. Soc. {\bf 380}, 1255 (2007), arXiv:0707.0521.

\bibitem{art:Frith2003}
W.~J. Frith {\em et~al.},
\newblock Mon. Not. Roy. Astr. Soc. {\bf 345}, 1049 (2003),
  arXiv:astro-ph/0302331.

\bibitem{art:Park2012}
C.~Park {\em et~al.},
\newblock Astrophys. J. {\bf 759}, L7 (2012), arXiv:1209.5659.

\bibitem{art:Dunkleyetal2009}
J.~Dunkley {\em et~al.},
\newblock Astrophys. J. Suppl. {\bf 180}, 306 (2009), arXiv:0803.0586.

\bibitem{art:Ade2013}
P.~A.~R. Ade {\em et~al.},
\newblock Planck 2013 results. {XXV}. {S}earches for cosmic strings and other
  topological defects,
\newblock arXiv:1303.5085, 2013.

\bibitem{art:Urrestilla2011}
J.~Urrestilla, N.~Bevis, M.~Hindmarsh, and M.~Kunz,
\newblock JCAP {\bf 12}, 021 (2011), arXiv:1108.2730.

\bibitem{art:Watkins2009}
R.~Watkins, H.~A. Feldman, and M.~J. Hudson,
\newblock Mon. Not. Roy. Astr. Soc. {\bf 392}, 743 (2009), arXiv:0809.4041.

\bibitem{art:Feldman2008}
H.~A. Feldman, M.~J. Hudson, and R.~Watkins,
\newblock Cosmic flows on 100 {M}pc/h scales,
\newblock arXiv:0805.1721, 2008.

\bibitem{art:Kashlinsky2011}
A.~Kashlinsky, F.~Atrio-Barandela, and H.~Ebeling,
\newblock Astrophys. J. {\bf 732}, 1 (2011), arXiv:1012.3214.

\bibitem{art:Kashlinsky2012}
A.~Kashlinsky, F.~Atrio-Barandela, and H.~Ebeling,
\newblock Measuring bulk motion of {X}-ray clusters via the kinematic
  {S}unyaev-{Z}eldovich effect: summarizing the ``dark flow" evidence and its
  implications,
\newblock arXiv:1202.0717, 2012.

\bibitem{art:Wiltshireetal2013}
D.~L. Wiltshire, P.~R. Smale, T.~Mattsson, and R.~Watkins,
\newblock Phys. Rev. D {\bf 88}, 083529 (2013), arXiv:1201.5371.

\end{thebibliography}

\end{document}